\documentclass{article}

\usepackage{arxiv}

\usepackage[utf8]{inputenc} 
\usepackage[T1]{fontenc}    
\usepackage{hyperref}       
\usepackage{url}            
\usepackage{booktabs}       
\usepackage{amsfonts}       
\usepackage{nicefrac}       
\usepackage{microtype}      
\usepackage{lipsum}		
\usepackage{amssymb}
\usepackage{amsmath}
\usepackage{graphicx}
\usepackage{xfrac}

\title{Estimating the Mutual Information between two Discrete, Asymmetric Variables with Limited Samples}

\author{
  Dami\'{a}n G. Hern\'{a}ndez \\
  Department of Medical Physics\\
  Centro At\'{o}mico Bariloche and Instituto Balseiro\\
  San Carlos de Bariloche, Argentina \\
   \And
 In\'{e}s Samengo \\
  Department of Medical Physics\\
  Centro At\'{o}mico Bariloche and Instituto Balseiro\\
  San Carlos de Bariloche, Argentina \\
}

\begin{document}
\maketitle

\begin{abstract}
Determining the strength of non-linear statistical dependencies between two variables is a crucial matter in many research fields. The established measure for quantifying such relations is the mutual information. However, estimating mutual information from limited samples is a challenging task. Since the mutual information is the difference of two entropies, the existing Bayesian estimators of entropy may be used to estimate information. This procedure, however, is still biased in the severely under-sampled regime. Here we propose an alternative estimator that is applicable to those cases in which the marginal distribution of one of the two variables---the one with minimal entropy---is well sampled. The other variable, as well as the joint and conditional distributions, can be severely undersampled. We obtain an estimator that presents very low bias, outperforming previous methods even when the sampled data contain few coincidences. As with other Bayesian estimators, our proposal focuses on the strength of the interaction between two discrete variables, without seeking to model the specific way in which the variables are related. A distinctive property of our method is that the main data statistics determining the amount of mutual information is the inhomogeneity of the conditional distribution of the low-entropy variable in those states (typically few) in which the large-entropy variable registers coincidences.
\end{abstract}

\keywords{bayesian estimation, mutual information, bias, sampling}

\section{Introduction}

Inferring the statistical dependencies between two variables from a few measured samples is an ubiquitous task in many areas of study. Variables are often linked through non-linear relations, which contain stochastic components. The standard measure employed to quantify the amount of dependency is the mutual information, defined as the reduction in entropy of one of the variables when conditioning the other variable \cite{shannon1948mathematical, cover2012elements}. If the states of the joint distribution are well-sampled, the joint probabilities can be estimated by the observed frequencies, yielding the maximum-likelihood estimator of mutual information. However, this procedure on average over-estimates the mutual information \cite{panzeri1996analytical, samengo2002estimating, paninski2003estimation}, so that independent variables may appear to be correlated, especially when the number of samples is small.  


The search for an estimator of mutual information that remains approximately unbiased even with small data samples is an open field of research \cite{kraskov2004estimating, montemurro2007tight, archer2013bayesian, kolchinsky2017estimating, belghazi2018mine, safaai2018information}. Here we focus on discrete variables, and assume it is not possible to overcome the scarceness of samples by grouping elements that are close according to some metric. In addition to corrections that only work in the limit of large samples \cite{strong1998entropy}, the state of the art for this problem corresponds to quasi-Bayesian methods that estimate mutual information indirectly through measures of the entropies of the involved variables \cite{archer2013bayesian, nemenman2004entropy, archer2014bayesian}. These approaches have the drawback of not being strictly Bayesian, since the linear combination of two or more Bayesian estimates of entropies does not, in general, yield a Bayesian estimator of the combination of entropies \cite{archer2013bayesian}. The concern is not so much to remain within theoretical Bayesian purity, but rather, to avoid frameworks that may be unnecessarily biased, or where negative estimates of information may arise. 

Here we propose a new method for estimating mutual information that is valid in the specific case in which there is an asymmetry between the two variables: One of them has a large number of effective states, and the other only a few. No hypotheses are made about the probability distribution of the large-entropy variable, but the marginal distribution of the low-entropy variable is assumed to be well sampled. The prior is chosen so as to accurately represent the amount of dispersion of the conditional distribution of the low-entropy variable around its marginal distribution. The main finding is that our estimator has very low bias, even in the severely under-sampled regime where there are few coincidences, that is, when a given state of the large-entropy variable is only seldom sampled more than once. The key data statistics that determine the estimated information is the inhomogeneity of the distribution of the low-entropy variable in those states of the high-entropy variable where two or more samples are observed. In addition to providing a practical algorithm to estimate mutual information, our approach sheds light on the way in which just a few samples reveal those specific properties of the underlying joint probability distribution that determine the amount of mutual information.

\section{Bayesian approaches to the estimation of entropies}

We seek a low-bias estimate of the mutual information between two discrete variables.   Let $X$ be a random variable with a large number $k_x$ of effective states $\{x_1, \dots, x_{k_x}\}$ with probabilities $q_x$, and $Y$ be a variable that varies in a small set $y\in \{y_1, \dots, y_{k_y}\}$, with $k_y \ll k_x$ . Given the conditional probabilities $q_{y|x}$, the marginal and joint probabilities are $q_y = \sum_x q_x \ q_{y|x}$ and $q_{xy} = q_x \ q_{y|x}$, respectively. The entropy $H(Y)$ is
\begin{equation}
\displaystyle H(Y)=-\sum_y q_y \log q_y,
\label{e00:entropydef}
\end{equation}
and can be interpreted as the average number of well-chosen yes/no questions required to guess the sampled value of $Y$ (when using a logarithm of base two). The conditional entropy $H(Y|X)$ is the average uncertainty of the variable $Y$ once $X$ is known, 
\begin{equation}
\displaystyle H(Y|X)=\sum_x q_x \left[-\sum_y q_{y|x} \log q_{y|x} \right]=\sum_x q_x H(Y|x).
\label{e00c:condentropydef}
\end{equation}
The mutual information is the reduction in uncertainty of one variable once we know the other \cite{cover2012elements} 
\begin{equation}
I(X,Y)=H(X)+H(Y)-H(X,Y)=H(Y)-H(Y|X).
\label{e01:infodef}
\end{equation}
Our aim is to estimate $I(X, Y)$ when $Y$ is well sampled, but $X$ is severely undersampled, in particular, when the sampled data contain few coincidences in $X$. Hence, for most values $x$, the number of samples $n_x$ is too small to estimate the conditional probability $q_{y|x}$ from the frequencies $n_{xy}/ n_x$. In fact, when $n_x \sim \mathcal{O}(1)$, the maximum likelihood estimator typically underestimates $H(Y|x)$ severely \cite{paninski2003estimation}, and consequently leads to an overestimation of $I(X,Y)$.

One possibility is to estimate $H(X), H(Y)$ and $H(X,Y)$ using a Bayesian estimator, and then plug the obtained values in Eq.~\ref{e01:infodef} to estimate the mutual information. We now discuss previous approaches to Bayesian estimators for entropy, to later analyze the case of information. For definiteness, we focus on $H(X)$, but the same logic applies to $H(Y)$, or $H(X, Y)$.

The Bayesian estimator is the expected value of $H(\mathbf{q}_x | \mathbf{n})$, where $\mathbf{q}_x$ are unknown probabilities $q_{x_1}, \dots, q_{x_k}$, and $\mathbf{n}$ represents the number of sampled data $\mathbf{n} = (n_1, \dots, n_k)$ obtained in each state. That is,
\begin{equation}
\begin{array}{c}
\langle H | \mathbf{n} \rangle  =  \int {\rm d} \mathbf{q} \ H(\mathbf{q}) \ p(\mathbf{q} | \mathbf{n}) 
= [p(\mathbf{n})]^{-1} \int {\rm d}\mathbf{q} \ H(\mathbf{q}) \ p(\mathbf{n} | \mathbf{q}) \ p(\mathbf{q}).
\end{array}
\label{e100:bayes}
\end{equation}
Since $p(\mathbf{n} | \mathbf{q})$ is the multinomial distribution
\begin{equation}
p(\mathbf{n} | \mathbf{q}) = N! \prod_{x} \frac{q_x^{n_x}}{n_x!},
\label{e105:multinomial}
\end{equation}
and since the normalization constant $p(\mathbf{n})$ can be calculated from the integral 
\begin{equation}
p(\mathbf{n}) = \int {\rm d}\mathbf{q} \  p(\mathbf{n} | \mathbf{q}) \ p(\mathbf{q}),
\label{e106:normalizacion}
\end{equation}
the entire gist of the Bayesian approach is to find an adequate prior $p(\mathbf{q})$ to plug into Eqs.~\ref{e100:bayes},~\ref{e105:multinomial} and \ref{e106:normalizacion}. For the sake of analytical tractability, $p(\mathbf{q})$ is often decomposed into a weighted combination of distributions $p(\mathbf{q} | \beta)$ that can be easily integrated, each tagged by one or or a few parameters, here generically called $\beta$, that vary within a certain domain,
\begin{equation}
p(\mathbf{q}) = \int {\rm d}\beta \ p(\beta) \ p(\mathbf{q} | \beta).
\label{e101:beta}
\end{equation}
The decomposition requires to introduce a prior $p(\beta)$. Hence, the former search for an adequate prior $p(\mathbf{q})$ is now replaced by the search for an adequate prior $p(\beta)$. The replacement implies an assumption and also a simplification. The family of priors that can be generated by Eq.~\ref{e101:beta} does not encompass the entire space of possible priors. The decomposition relies on the assumption that the remaining family is still rich enough to make good inference about the quantity of interest, in this case, the entropy. The simplification stems from the fact that the search for $p(\beta)$ is more restricted than the search for $p(\mathbf{q})$, because the space of possible alternatives is smaller (the dimensionality of $\mathbf{q}$ is typically high, whereas the one of $\beta$ is low).  Two popular proposals of Bayesian estimators for entropies are NSB \cite{nemenman2004entropy} and PYM \cite{archer2014bayesian}. In NSB, the functions $p(\mathbf{q} | \beta)$ are Dirichlet distributions, in which $\beta$ takes the role of a concentration parameter. In PYM, these functions are Pitman-Yor processes, and $\beta$ stands for two parameters: one accounting for the concentration, and the other for the so-called discount. In both cases, the Bayesian machinery implies
\begin{equation}
\langle H | \mathbf{n} \rangle =  \frac{1}{p(\mathbf{n})} \int {\rm d}\beta \ p(\beta) \ W(\beta | \mathbf{n}),
\label{e107:machinery}
\end{equation}
where $W(\beta|\mathbf{n})$ is the weight of each $\beta$ in the estimation of the expected entropy
\begin{equation}
W(\beta | \mathbf{n}) =  \int {\rm d}\mathbf{q} \ H(\mathbf{q}) \ p(\mathbf{n} | \mathbf{q}) \ p(\mathbf{q} | \beta). \label{e107:peso}
\end{equation}
When choosing the family of functions $p(\mathbf{q} | \beta)$, it is convenient to select them in such a way that the weight $W(\beta|\mathbf{n})$ can be solved analytically. However, this is not the only requirement. In order to calculate the integral in $\beta$, the prior $p(\beta)$ also plays a role. The decomposition of Eq.~\ref{e101:beta} becomes most useful when the arbitrariness in the choice of $p(\beta)$ is less serious than the arbitrariness in the choice of $p(\mathbf{q})$. This assumption is justified when $W(\beta | \mathbf{n})$ is peaked around a specific $\beta$ value, so that in practice, the shape of $p(\beta)$ hardly has an effect. In these cases, a narrow range of relevant $\beta$ values is selected by the sampled data, and all assumptions about the prior probability outside this range play a minor role. For the choices of the families $p(\mathbf{q}|\beta)$ proposed by NSB and PYM, $W(\beta | \mathbf{n})$ can be calculated analytically, and one can verify that indeed, a few coincidences in the data suffice for a peak to develop. In both cases, the selected $\beta$ is one for which $p(\mathbf{q} | \beta)$ favours a range of $\mathbf{q}$ values that are compatible with the measured data (as assessed by $p(\mathbf{n}|\mathbf{q})$), and also produce non-negligible entropies (Eq.~\ref{e107:peso}).

When the chosen Bayesian estimates of the entropies are plugged into Eq.~\ref{e01:infodef} to obtain an estimate of the information, each term is dominated by its own preferred $\beta$. Since the different entropies are estimated independently, the $\beta$ values selected by the data to dominate the priors $p(q_{x})$ and $p(q_y)$ need not be compatible with the ones dominating the priors of the joint or the conditional distributions. As a consequence, the estimation of the mutual information is no longer Bayesian, and can suffer from theoretical issues, as for example, yield a negative estimate \cite{archer2013bayesian}.

A first alternative would be to consider an integrable prior containing a single $\beta$ for the joint probability distribution $q_{xy}$, and then replace $H$ by $I$ in the equations above, to calculate $\langle I \rangle$. This procedure was tested by Archer et al.  \cite{archer2013bayesian}, and the results were only good when the collection of $q_{xy}$ values governing the data were well described by a distribution that was contained in the family of proposed priors $p(\mathbf{q} | \beta)$. The authors concluded that mixtures of Dirichlet priors do not provide a flexible enough family of priors for highly-structured joint distributions, at least for the purpose of estimating mutual information.

To make progress, we note that $I(X, Y)$ can be written as
\begin{equation}
I(X; Y) = \sum_x q_x \,  \sum_y q_{y|x}\,\log\frac{q_{y|x}}{q_y} = 
\sum_x q_x\, D_{\text{KL}}(\mathbf{q}_{y|x}|| \mathbf{q}_y),
\label{eq22:thm1}
\end{equation}
where $\mathbf{q}_{y|x}$ and $\mathbf{q}_y$ stand for the $k_y$-dimensional vectors $(q_{y_1|x}, \dots, q_{y_{k_y} | x})$ and $(q_{y_1}, \dots, q_{y_{k_y}})$, and $D_{\text{KL}}$ represents the Kullback-Leibler divergence. The average divergence between $\mathbf{q}_{y|x}$ and $\mathbf{q}_y$ captures a notion of spread. Therefore, the mutual information is sensitive not so much to the value of the probabilities $q_{y|x}$, but rather, to their degree of scatter around the marginal $q_y$. The parameters controlling the prior should hence be selected in order to match the width of the distribution of $q_{y|x}$ values, and not so much each probability. With this intuition in mind, in this paper we put forward a new prior for the whole ensemble of conditional probabilities $q_{y|x}$ obtained for different $x$ values. In this prior, the parameter $\beta$ controls the spread of the conditionals $q_{y|x}$ around the marginal $q_y$. 


\section{Prior distribution for conditional entropies}

Our approach is valid when the total number of samples $N$ is at least of the order of magnitude of $\sqrt{e^{H(X)}}$, since in this regime, some of the $x$ states are expected to be sampled more than once \cite{ma1981,nemenman2011coincidences}. In addition, the marginal distribution $q_y$ must be well sampled. This regime is typically achieved when $X$ has a much larger set of available states than $Y$. In this case, the maximum likelihood estimators $\hat{q}_y$ of the marginal probabilities $q_y$ can be assumed to be accurate, that is,
\begin{equation}
\hat{q}_y = \frac{n_y}{N} \approx q_y, \ \ \ \forall \ y.
\end{equation}

In this paper, we put forward a Dirichlet prior distribution centered at $\hat{q}_y$, that is,
\begin{eqnarray}
p(\{\mathbf{q}_{y|x}\}| \beta) &=&   \Gamma(\beta)^{k_x} \prod_{xy} \frac{\left.q_{y|x}\right.^{\beta \hat{q}_y - 1}}{\Gamma(\beta \hat{q}_y)} \nonumber \\
&\propto& \frac{\exp\left[-\beta \sum_x D_{\rm KL}(\hat{\mathbf{q}_y}||\mathbf{q}_{y|x}) \right] }{\prod_{xy} q_{y|x}}, \label{e107:nuestraprior} 
\end{eqnarray}
where $\{\mathbf{q}_{y|x}\}$ contains the $k_x$ conditional probabilities $\mathbf{q}_{y|x}$ corresponding to different $x$ values. Large $\beta$ values select conditional probabilities close to $\hat{q}_y$, while small values imply a large spread, that pushes the selection towards the border of the $k_y$-simplex.

For the moment, for simplicity we work with a prior $p(\{\mathbf{q}_{y|x}\})$ defined on the conditional probabilities $q_{y|x}$, and make no effort to model the prior probability of the vector $\mathbf{q}_x$. In practice, we estimate the values of $q_x$ with the maximum likelihood estimator $\hat{q}_x = n_x / N$. Since $X$ is assumed to be severely undersampled, this is a poor procedure to estimate $q_x$. Still, the effect on the mutual information turns out to be negligible, since the only role of $q_x$ in Eq.~\ref{eq22:thm1} is to weigh each of the Kullback-Leibler divergences appearing in the average. If $k_x$ is large, each $D_{KL}$ value will appear in several terms of the sum, rendering the individual value of the accompanying $q_x$ irrelevant, only the sum of them matters. In Sect.~\ref{sec:inferringqx}, we tackle the full problem of making Bayesian inference both in $\mathbf{q}_x$ and $\{\mathbf{q}_{y|x}\}$.

The choice of prior of Eq.~\ref{e107:nuestraprior} is inspired in three facts. First, $\beta$ captures the spread of $q_{y|x}$ around $q_y$, as implied by the Kullback-Leibler divergence in Eq.~\ref{e107:nuestraprior}. Admittedly, this divergence is not exactly the one governing the mutual information (Eq.~\ref{eq22:thm1}), since $\mathbf{q}_{y|x}$ and $\mathbf{q}_y$ are interchanged. Yet, it is still a measure of spread. The exchange, as well as the denominator in Eq.~\ref{e107:nuestraprior}, were introduced for the sake of the second fact, namely, analytical tractability. The third fact regards the emergence of a single relevant $\beta$ when the sampled data begin to register coincidences. If we follow the Bayesian rationale of the previous section, now replacing the entropy by the mutual information, we can again define a weight $W(\beta | \mathbf{n})$ for the parameter $\beta$
\begin{eqnarray}
W(\beta | \mathbf{n}) &=& \int \{{\rm d}\mathbf{q}_{y|x}\} \ I(\hat{\mathbf{q}}_x, \{\mathbf{q}_{y|x}\}) \ p(\mathbf{n} | \hat{\mathbf{q}_x}, \{\mathbf{q}_{y|x}\}) \ p(\{\mathbf{q}_{y|x}\} | \beta) \nonumber \\
&=& p(\beta|\mathbf{n}) \ F(\beta, \mathbf{n}) \nonumber,
\end{eqnarray}
where $F(\beta, \mathbf{n})$ can be obtained analytically, and is a well behaved function of its arguments, whereas
\begin{eqnarray}
p(\beta | \mathbf{n}) &=& \frac{p(\beta) \ p(\mathbf{n}|\beta)}{p(\mathbf{n})} =  \frac{p(\beta)}{p(\mathbf{n})} \ \int \left\{{\rm d}\mathbf{q}_{y|x}\right\} \ p(\mathbf{n}|\hat{\mathbf{q}}_x, \{\mathbf{q}_{y|x}\}) \ p(\{\mathbf{q}_{y|x}\}|\beta) \nonumber \\
&=& \frac{p(\beta)}{p(\mathbf{n})} \prod_x \frac{\Gamma(\beta)}{\Gamma(n_{x}+\beta)} \prod_{y=1}^{k_y} \frac{\Gamma(n_{xy}+\beta\hat{q}_y)}{\Gamma(\beta\hat{q}_y)}. \label{e108:pbetan}
\end{eqnarray}
For each $x$, the vector $\mathbf{q}_{y|x}$ varies in a $k_y$-dimensional simplex. For $p(\mathbf{n}|\hat{\mathbf{q}}_x, \{\mathbf{q}_{y|x}\})$ we take the multinomial
\begin{equation}
p(\mathbf{n}|\hat{\mathbf{q}}_x, \{\mathbf{q}_{y|x}\}) = N! \prod_{xy} \frac{[\hat{q}_x \ q_{y|x}]^{n_{xy}}}{n_{xy}!}.
\end{equation}
The important point here, is that the ratio of the Gamma functions of Eq.~\ref{e108:pbetan} develops a peak in $\beta$ as soon as the collected data register a few coincidences in $x$. Hence, with few samples, the prior proposed in Eq.~\ref{e107:nuestraprior} renders the choice of $p(\beta)$ inconsequential.

Assuming that the marginal probability of $Y$ is well-sampled, the entropy $H(Y)$ is well approximated by the maximum-likelihood estimator $\hat{H}(Y) = -\sum_y (n_y/N) \ \log(n_y/N)$. For each $\beta$, the expected posterior information can be calculated analytically,
\begin{equation}
\langle I|\mathbf{n} \rangle (\beta) =\hat{H}(Y) -\sum_x \frac{n_x}{N} \ \left[\psi_0(\beta + n_x + 1) -\sum_y \frac{\beta \hat{q_y} + n_{xy}}{\beta + n_x} \psi_0(\beta \hat{q_y} + n_{xy} + 1)  \right],
\label{e18:postinfok}
\end{equation}
where $\psi_0$ is the digamma function. When the system is well sampled, $n_{xy}\gg 1$, so the effect of $\beta$ becomes negligible, the Digamma functions tend to logarithms, and the frequencies match the probabilities. In this limit, Eq.~\ref{e18:postinfok} coincides with the maximal likelihood estimator, which is consistent. The rest of the paper focuses on the case in which the marginal probability of $X$ is severely undersampled.


\section{A closer look on the case of a symmetric and binary $Y$-variable}

\begin{figure}[ht]
  \centering
  \includegraphics[width=0.65\linewidth]{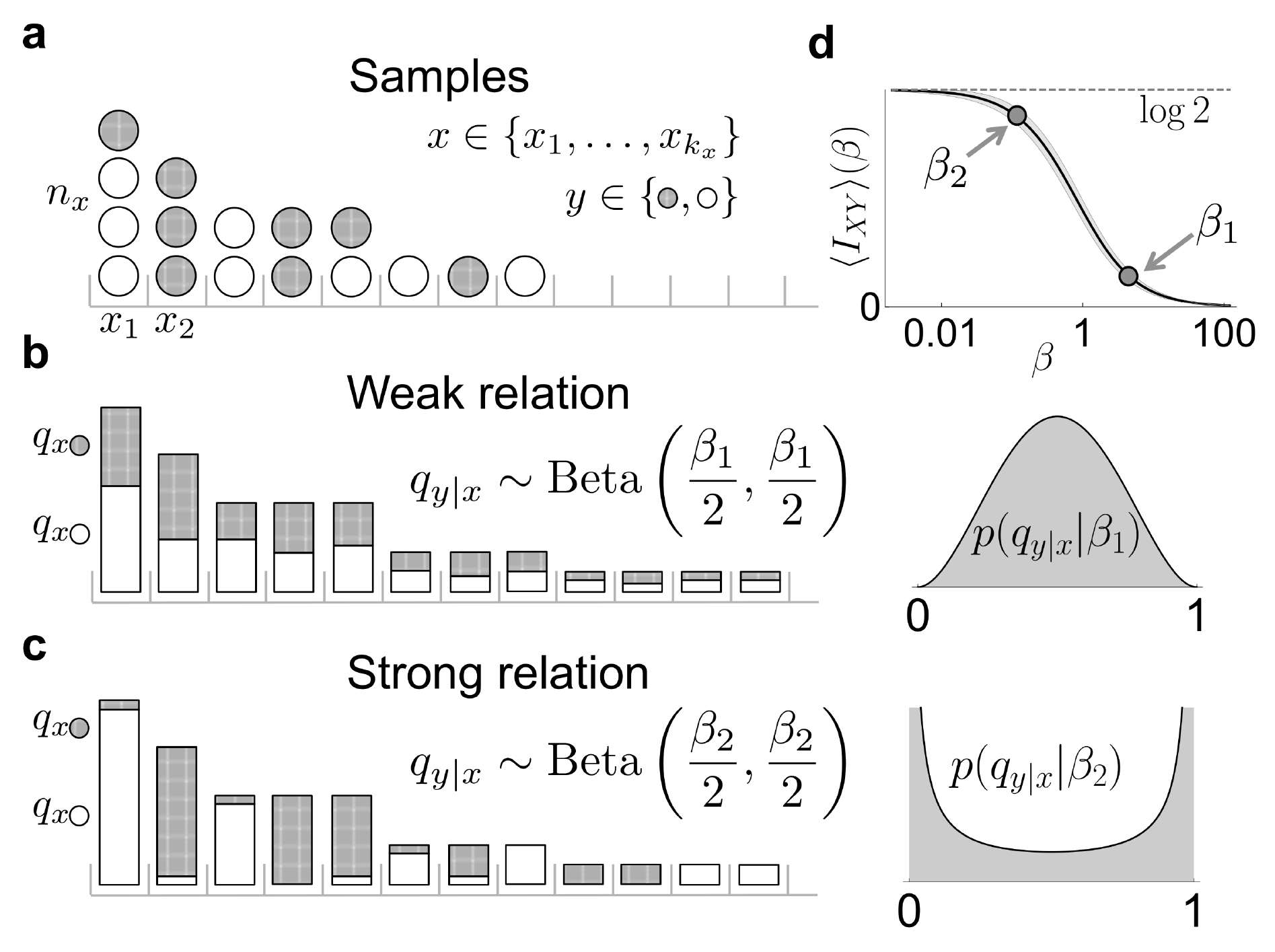}
  \caption{A scheme of our method to estimate the mutual information between two variables $X$ and $Y$. {\bf a}: We collect a few samples of a variable $x$ with a large number of effective states $x_1, x_2, \dots$, each sample characterized by a binary variable $y$ (the two values represented in white and gray). We consider different hypotheses about the strength with which the probability of each $y$-value varies with $x$. {\bf b}: One possibility is that the conditional probability of each of the two $y$-values hardly varies with $x$. This situation is modeled by assuming that the different $q_{y|x}$ are random variables governed by a Beta distribution with a large hyper-parameter $\beta_1$. {\bf c}: On the other hand, the conditional probability $q_{y|x}$ could vary strongly with $x$. This situation is modeled by a Beta distribution with a small hyper-parameter $\beta_2$. {\bf d}: As $\beta$ varies, so does the prior mutual information (Eq.~\ref{e03:priorinfo}). If the distribution $p(\mathbf{q}|\beta)$ is sampled repeatedly for a fixed $\beta$, the prior information $\langle I(\mathbf{q}) \rangle$ may fluctuate from sample to sample. The shaded area around the solid line illustrates such fluctuations when $k_x = 50$.}
\label{f01:scheme}
\end{figure}

In this section, for simplicity we take $q_{y=0}=q_{y=1} = \sfrac{1}{2}$, such that $H(Y)=\log 2$ nats.  In this case, the Dirichlet prior of Eq.~\ref{e107:nuestraprior} becomes a Beta distribution
\begin{equation}
\displaystyle p(q_{1|x}|\beta) = \frac{\Gamma(\beta)}{\Gamma(\beta/2)^2} \left[q_{1|x}(1-q_{1|x})\right]^{\beta/2-1}.
\label{e02:condq}
\end{equation}
Large values of $\beta$ mostly select conditional probabilities $q_{y|x}$ close to $1/2$. If all conditional probabilities are similar, and similar to the marginal, the mutual information is low, since the probability of sampling a specific $y$ value hardly depends on $x$. Instead, small values of $\beta$ produce conditional probabilities $q_{y|x}$ around the borders ($q_{y|.}\sim 0$ or $q_{y|.}\sim 1$). In this case, $q_{y|x}$ is strongly dependent on $x$ (see Fig.~\ref{f01:scheme} {\bf b}), so the mutual information is large. The expected prior mutual information $\langle I(\beta) \rangle$ can be calculated using the analytical approach developed by \cite{wolpert1995estimating, archer2014bayesian},
\begin{equation}
\begin{array}{rl}
\langle I \rangle (\beta) &= \log 2 -\psi_0(\beta+1)+\psi_0(\beta/2+1).
\end{array}
\label{e03:priorinfo}
\end{equation}
The prior information is a slowly-varying function of the order of magnitude of $\beta$, namely of $\log\beta$. Therefore, if a uniform prior  in information is desired, it suffices to choose a prior on $\log\beta$ such that $p(\log\beta) \propto |\partial_{\log\beta}\langle I \rangle (\beta)|$,
\begin{equation}
p(\log\beta)=\displaystyle\frac{\beta/2}{\log 2}\, |2 \psi_1(\beta+1)-\psi_1(\beta/2+1)|.
\label{e04:priorb}
\end{equation}

When $k_y = 2$, the expected posterior information (Eq.~\ref{e18:postinfok}) becomes
\begin{equation}
\langle I|\mathbf{n} \rangle (\beta)= \hat{H}(Y)-\sum_x \frac{n_x}{N}\left[ \psi_0(\beta+n_x+1)  -\sum_{y \in \{0, 1\}} \left(\frac{\beta/2+n_{xy}}{\beta+n_x}\right)\psi_0(\beta/2+n_{xy}+1)\right].
\label{e06:postinfo}
\end{equation}

The marginal likelihood of the data given $\beta$ is also analytically tractable. The likelihood is binomial for each $x$, so
\begin{eqnarray}
p(\mathbf{n}|\beta) &=& \prod_x \int_0^1 {\rm d}q_{1|x}\,\,p(n_{x1}, n_{x0}|q_{1|x})\,p(q_{1|x}|\beta) \nonumber \\
&\propto& \prod_x \frac{\Gamma(n_{x1}+\beta/2)\Gamma(n_{x0}+\beta/2)\Gamma(\beta)}{\Gamma(n_{x}+\beta){\Gamma(\beta/2)}^2}.
\label{e05:evidenceb}
\end{eqnarray}

The posterior for $\beta$ can be obtained by adding a prior $p(\beta)$, as $p(\beta|\mathbf{n}) \propto p(\mathbf{n}|\beta)p(\beta)$. The role of the prior becomes relevant when the number of coincidences is too low for the posterior to develop a peak (see below). 


In order to gain intuition about the statistical dependence between variables with few samples, we here highlight the specific aspects of the data that influence the estimator of Eq.~\ref{e06:postinfo}. Grouping together the terms of Eq.~(\ref{e05:evidenceb}) that are equal, the marginal likelihood can be rewritten in terms of the multiplicities $m_{nn'}$, that is, the number of states $x$ with specific occurrences $\{n_{x1}=n,\,n_{x0}=n'\}$ or $\{n_{x1}=n',\,n_{x0}=n\}$,
\begin{equation}
\begin{array}{rl}
\log p(\mathbf{n}|\beta)&\displaystyle=\sum_{n\geq n'} m_{nn'}\log\left[\frac{\Gamma(n+\beta/2)\Gamma(n'+\beta/2)\Gamma(\beta)}{\Gamma(n+n'+\beta)\Gamma(\beta/2)^2}\right]\\\\
&\displaystyle=\sum_{n\geq n'} m_{nn'}\log p_{nn'}(\beta),
\end{array}
\label{e07:multipb}
\end{equation}
where 
\begin{equation}
\begin{array}{rl}
p_{10}(\beta)&\displaystyle=\frac{\beta/2}{\beta}=\frac{1}{2}\\\\
p_{11}(\beta)&\displaystyle=\frac{(\beta/2)^2}{\beta(\beta+1)}=\frac{\beta}{4(\beta+1)}\\\\
p_{20}(\beta)&\displaystyle=\frac{(\beta/2)(\beta/2+1)}{\beta(\beta+1)}=\frac{(\beta/2+1)}{2(\beta+1)}\\
\dots \\
p_{nn'}(\beta)&\displaystyle=\frac{(\beta/2)(\beta/2+1)\dots(\beta/2+n-1)\,(\beta/2)(\beta/2+1)\dots(\beta/2+n'-1)}{\beta(\beta+1)\dots(\beta+n+n'-1)}.
\end{array}
\label{e08:pnnb}
\end{equation}
The posterior for $\beta$ is independent from states $x$ with just a single count, as $p_{10}(\beta)=\text{constant}$. Only states $x$ with coincidences matter. In order to see how the sampled data favor a particular $\beta$, we search for the $\beta$ value that maximizes $\log p(\mathbf{n}|\beta)$ in the particular case where at most two samples coincide on the same $x$, obtaining
\begin{equation}
\frac{\partial}{\partial\beta}\log p(\mathbf{n}|\beta)=\frac{m_{11}}{\beta}+\frac{m_{20}}{\beta+2}-\frac{m_{11}+m_{20}}{\beta+1} = 0.
\label{e09:sol2coutns}
\end{equation}
Denoting the fraction of $2$-count states that have one count for each $y$ value as $f_{11}=m_{11}/(m_{11}+m_{20})$, Eq.~\ref{e09:sol2coutns} implies that $\beta \rightarrow \infty$ if $f_{11}\geq 1/2$, and $\beta=f_{11}/(1/2- f_{11})$, otherwise. If the $y$-values are independent of $x$, we expect $f_{11}\sim 1/2$. This case corresponds to a large $\beta$ and, consequently, to a low information. On the other side, for small $f_{11}$, the parameter $\beta$ is also small and the information grows. 

In Eq.~\ref{e09:sol2coutns}, the data only intervene through $m_{11}$ and $m_{20}$, which characterize the degree of asymmetry of the $y$ values throughout the different $x$ states. This asymmetry, hence, constitutes a sufficient statistics for $\beta$. If a prior $p(\beta)$ is included, the $\beta$ that maximizes the posterior $p(\beta | \mathbf{n})$ may shift, but the effect becomes negligible as the number of coincidences grows.

We now discuss the role of the selected $\beta$ in the estimation of information, Eq.~(\ref{e06:postinfo}), focusing on the conditional entropy $\langle H_{Y|X}\rangle(\beta)$. First, in terms of the multiplicities, the conditional entropy can be rewritten as
\begin{equation}
\displaystyle \langle H_{Y|X}\rangle(\beta)= \sum_k f_k \sum_{n+n'=k} f_{nn'} H_{nn'}(\beta)
\label{e10:hydadoxb}
\end{equation}
where $f_k$ is the fraction of the $N$ samples that fall in states $x$ with $k$ counts, and $f_{nn'}$ is the fraction of all states $x$ with $n+n'$ counts that have $n$ for one $y$-value (whichever) and $n'$ for the other. Finally, $H_{nn'}(\beta)$ is the estimation of the entropy of a binary variable after $\{n,n'\}$ samples,
\begin{equation}
\displaystyle H_{nn'}(\beta)=  \psi_0(n+n'+\beta+1) -\frac{(n+\beta/2)\psi_0(n+\beta/2+1)+(n'+\beta/2)\psi_0(n'+\beta/2+1)}{n+n'+\beta}.\label{e11:hnnb}
\end{equation}
A priori, $\langle I \rangle(\beta)=\log 2-H_{00}(\beta)$, as in Fig.~\ref{f01:scheme}{\bf d}. Surprisingly, from the property $\psi_0(z+1)=\psi_0(z)+1/z$, it turns out that $H_{00}=H_{10}$ (in fact, $H_{nn}=H_{(n+1)n}$). Hence, if only a single count breaks the symmetry between the two $y$ values, there is no effect on the conditional entropy. This is a reasonable result, since a single extra count is no evidence of an imbalance between the underlying conditional probabilities, it is just the natural consequence of comparing the counts falling on an even number of states (2) when taking an odd number of samples. Expanding the first terms for the conditional entropy,
\begin{equation}
\begin{array}{l}
\displaystyle \langle H_{Y|X}\rangle= f_1 H_{00}(\beta)+f_2 f_{11} H_{11}(\beta)+f_2 f_{20} H_{20}(\beta)+\dots
\end{array}
\label{e12:hydadoxbaprox}
\end{equation}
In the severely under-sampled regime, these first terms are the most important ones. Typically, $f_1$ takes most of the weight, and Eq.~(\ref{e12:hydadoxbaprox}) implies that the estimation is close to the prior $H_{00}$ evaluated in the value of $\beta$ that maximizes the marginal likelihood (or the posterior).

Finally, we mention that when dealing with few samples, it is important to have not just a good estimate of the mutual information, but also, a confidence interval. Even a small information may be relevant, if the evidence attests that it is strictly above zero. The theory developed here also allows us to estimate the posterior variance of the mutual information, as shown in the Appendix. The variance (Eq.~\ref{e:errorih}) is shown to be inversely proportional to the number of states $k_x$, thereby implying that our method benefits from a large number of available states $X$, even if undersampled.


\begin{figure}
  \centering
  \includegraphics[width=\linewidth]{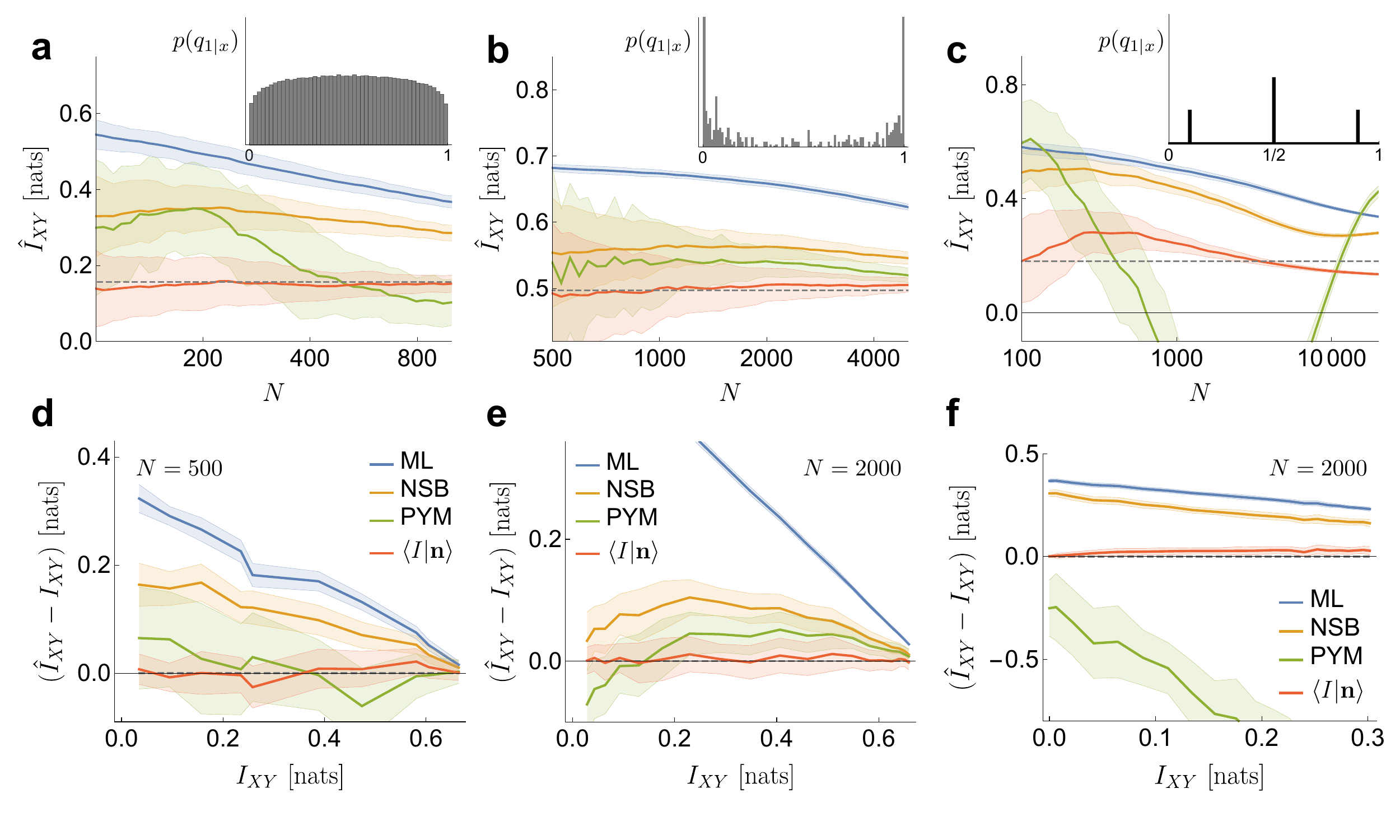}
  \caption{Comparison of the performance of four different estimators for $I_{XY}$: maximum likelihood estimator (ML), NSB estimator used in the limit of infinite states, PYM estimator, and our estimator $\langle I|\mathbf{n} \rangle (\beta)$ (Eq.~\ref{e06:postinfo}) calculated with the $\beta$ that maximizes the marginal likelihood $p(\mathbf{n} | \beta)$ (Eq.~\ref{e05:evidenceb}). The curves represent the average over $50$ different data sets $\mathbf{n}$, with the standard deviation displayed as a colored area around the mean. {\bf a}: Estimates of mutual information as a function of the total number of samples $N$, when the values of $q_{1|x}$ are generated under the hypothesis of our method (Eq.~\ref{e02:condq}). We sample once the marginal probabilities $q_x\sim \text{PYM}(\alpha=50, d=0.55)$, as well as the conditionals $q_{y|x}\sim \text{Beta}(\beta/2,\beta/2)$ with $\beta=2.3$. The effective size of the system is $\exp(H_{XY})\simeq 800$. The exact value of $I_{XY}$ is shown as a horizontal dashed line. {\bf b}: Estimates of mutual information, for data sets where the conditional probabilities have spherical symmetry. $X$, a binary variable of dimension $12$, corresponds to the presence of $12$ delta functions equally spaced in a sphere ($q_x = 2^{-12}$, for all $x$). We generate the conditional probabilities such that they are invariant under rotations of the sphere, namely $q_{y|x}=q_{y|\text{R}(x)}$, being R a rotation. To this aim, we set $q_{y|x}$ as a sigmoid function of a combination of frequency components ($\pi_0-\pi_1-\pi_2$) of the spherical spectrum \cite{kazhdan2003rotation}. The effective size of the system is $\exp(H_{XY})\simeq 5000$. {\bf c}: Estimates of mutual information, for a conditional distribution far away from our hypotheses. The $x$ states are generated as Bernoulli ($p=0.05$) binary vectors of dimension $D=40$, while the conditional probabilities depend on the parity of the sum of the components of the vector. When the sum is even we set $q_{y|x}=1/2$, and when is odd, $q_{y|x}$ is generated by sampling a mixture of two deltas of equal weight $q_{y|x}\sim [\delta(q-q_0)+\delta(q-1+q_0)]/2$ with $q_0=0.1$. The resulting distribution of $q_{y|x}$-values contains 3 peaks, and therefore, cannot be described with a Dirichlet distribution. The effective size of the system is $\exp(H_{XY})\simeq 4000$. {\bf d}: Bias in the estimation as a function of the value of mutual information. Settings remain the same as in {\bf a}, but fixing $N=500$ and changing $\beta\in(0.04,\,14)$ in the conditional. {\bf e}: Bias in the estimation as a function of the value of mutual information. Settings as in {\bf b}, but fixing $N=2000$ and changing the gain of the sigmoid in the conditional. {\bf f}: Bias in the estimation as a function of the value of mutual information. Settings as in {\bf c}, but fixing $N=2000$ and changing $q_0\in(0.01,\,0.4)$ in the conditional.}
\label{f02:est}
\end{figure}

\section{Testing the estimator}
\label{testing}

We now analyze the performance of our estimator in three examples where the number of samples $N$ is below or in the order of the effective size of the system $\exp(H_{XY})$. In this regime, most observed $x$ states have very few samples. In each example, we define the probabilities $q_x$ and $q_{1|x}$ with three different criteria, giving rise to collections of probabilities that can be described with varying success by the prior proposed in this paper, Eq.~\ref{e02:condq}. Once the probabilities are defined, the true value $I_{XY}$ of the mutual information can be calculated, and compared to the one estimated by our method, as well as by three other estimators employed in the literature, in 50 different sets of samples  $\mathbf{n}$ of the measured data. As our estimator we use $\langle I|\mathbf{n}\rangle$ from Eq.~(\ref{e06:postinfo}) evaluated in the $\beta$ that maximizes the marginal likelihood $p(\mathbf{n}|\beta)$. We did not observe any improvement when integrating over the whole posterior $p(\beta|\mathbf{n})$ with the prior $p(\beta)$ of Eq.~\ref{e04:priorb}, except when $m_{20}$ or $m_{10}$ were of order 1. This fact implies the existence of a well-defined peak in the marginal likelihood.

In the first example (Fig.~\ref{f02:est}{\bf a, d}), the probabilities $q_x$ are obtained by sampling a Pitman-Yor distribution with concentration parameter $\alpha=50$ and tail parameter $d=0.55$. These values correspond to a PYM prior with a heavy tail. The conditional probabilities $q_{y|x}$ are defined by sampling a symmetric Beta distribution $q_{y|x}\sim \text{Beta}(\beta/2,\beta/2)$, as in Eq.~\ref{e02:condq}. In Fig.~\ref{f02:est}{\bf a}, we use $\beta=2.3$. Once the joint probability $q_{xy}$ is defined, 50 sets of samples $\mathbf{n}$ are generated. The effective size of the system is $\exp(H_{XY})\simeq 800$. We compare our estimator to maximum likelihood (ML), NSB and PYM when applied to $H_{X}$ and $H_{XY}$ (all methods coincide in the estimation of $H_Y$). Our estimator has a low bias, even when the number of samples per effective state is as low as $N/e^{H_{XY}}=0.15$. The variance is larger than ML, comparable to NSB and smaller than PYM. All the other methods (ML, NSB and to a lesser extend PYM) overestimate the mutual information. In Fig.~\ref{f02:est}{\bf d}, the performance of the estimators is also tested for different values of the exact mutual information $I_{XY}$, which we explore by varying $\beta\in(0.04,\,14)$. For each $\beta$, the conditional probabilities $q_{1|x}$ are sampled once. Each vector $\mathbf{n}$ contains $N = 500$ samples, and $\mathbf{n}$ is sampled $50$ times. Our estimates have very low bias, even as the mutual information goes to zero ---namely, for independent variables.

Secondly, we analyze an example where the statistical relation between $X$ and $Y$ is remarkably intricate (example inspired by \cite{shwartz2017opening}), which underscores the fact that making inference about the mutual information does not require inferences on the joint probability distribution. The variable $x$ is a binary vector of dimension $12$. Each component represents the presence or absence of one of a maximum of $12$ delta functions equally spaced on the surface of a sphere. There are $2^{12}$ possible $x$ vectors, and they are governed by a uniform prior probability: $q_x = 2^{-12}$. The conditional probabilities are generated in such a way that they be invariant under rotations of the sphere, that is, $q_{y|x}=q_{y|\text{R}(x)}$, where $R$ is a rotation. Using a spherical harmonic representation \cite{kazhdan2003rotation}, the frequency components $\pi_\ell(f(x))$ of the spherical spectrum are obtained, where $f(x)$ is the combination of deltas. The conditional probabilities $q_{y|x}$ are defined as a sigmoid function of $(\pi_0-\pi_1-\pi_2)$. The offset of the sigmoid is chosen such that $q_{y=1}\simeq 0.5$, and the gain such that $I_{XY}\simeq 0.5$ nats. In this example, and unlike the Dirichlet prior implied by our estimator, $p(q_{y|x})$ has some level of roughness (inset in Fig.~\ref{f02:est}{\bf b}), due to peaks coming from the invariant classes in $\{x_1, \dots, x_{2^{12}}\}$. Hence, the example does not truly fit into the hypothesis of our method. With these settings, the effective size of the system is $\exp(H_{XY})\simeq 5000$. Our estimator has little bias (Figs.~\ref{f02:est}{\bf b, e}), even with $N/e^{H_{XY}}=0.2$ samples per effective state. In this regime, around $\sim 80\%$ of the samples fall on $x$ states that occur only once ($f_1\simeq 0.8$), $\sim 19\%$ on states that occur twice and $\sim 1\%$ on states with $3$ counts, or maybe $4$. As mentioned above, in such cases, the value of $I_{XY}$ is very similar to the one that would be obtained by evaluating the prior information $\langle I | \mathbf{n} = \mathbf{0}, \beta \rangle$ of (Eq.~\ref{e03:priorinfo}) at the $\beta$ that maximizes the marginal likelihood $p(\mathbf{n}|\beta)$, which in turn is mainly determined by $f_{11}$. In Fig.~\ref{f02:est}{\bf e}, the estimator is tested with a fixed number of samples $N=2000$ for different values of the mutual information, which we explore by varying the gain of the sigmoid. The bias of the estimate is small in the entire range of mutual informations.

In the third place, we consider an example where the conditional probabilities are generated from a distribution that is poorly approximated by a Dirichlet prior. The conditional probabilities are sampled from three Dirac deltas, as $q_{y|x}\sim [0.5\,\delta(q-\sfrac{1}{2})+0.25\,\delta(q-q_0)+0.25\,\delta(q-1+q_0)]$, with $q_0=0.1$. The delta placed in $q=\sfrac{1}{2}$ could be approximated by a Dirichlet prior with a large $\beta$, while the other two deltas could be approximated by a small $\beta$, but there is no single value of $\beta$ that can approximate all three deltas at the same time. The $x$ states are generated as Bernoulli ($p=0.05$) binary vectors of dimension $D=40$, while the conditional probabilities $q_{1|x}$ depend on the parity of the sum of the components of the vector $x$. When the sum is even, we assign $q_{y|x}=\sfrac{1}{2}$, and when it is odd, we assign $q_{y|x}=q_0$ or $q_{y|x}=1-q_0$, both options with equal probability. Although in this case our method has some degree of bias, it still preserves a good performance in relation to the other approaches (see Fig.~\ref{f02:est}{\bf c}, {\bf f}). The marginal likelihood $p(\mathbf{n} | \beta)$ contains a single peak in an intermediate value of $\beta$, coinciding with none of the deltas in $p(q_{1|x})$, but still capturing the right value of the mutual information. As in the previous examples we also test the performance of the estimator for different values of the mutual information, varying in this case the value of $q_0$ (with $N=2000$). Our method performs acceptably for all values of mutual information. The other methods, instead, are challenged more severely, probably because a large fraction of the $x$ states have a very low probability, and are therefore difficult to sample. Those states, however, provide a crucial contribution to the relative weight of each of the three values of $q_{1|x}$.  PYM, in particular, sometimes produces a negative estimate for $I_{XY}$. 

\begin{figure}
  \centering
  \includegraphics[width=0.8\linewidth]{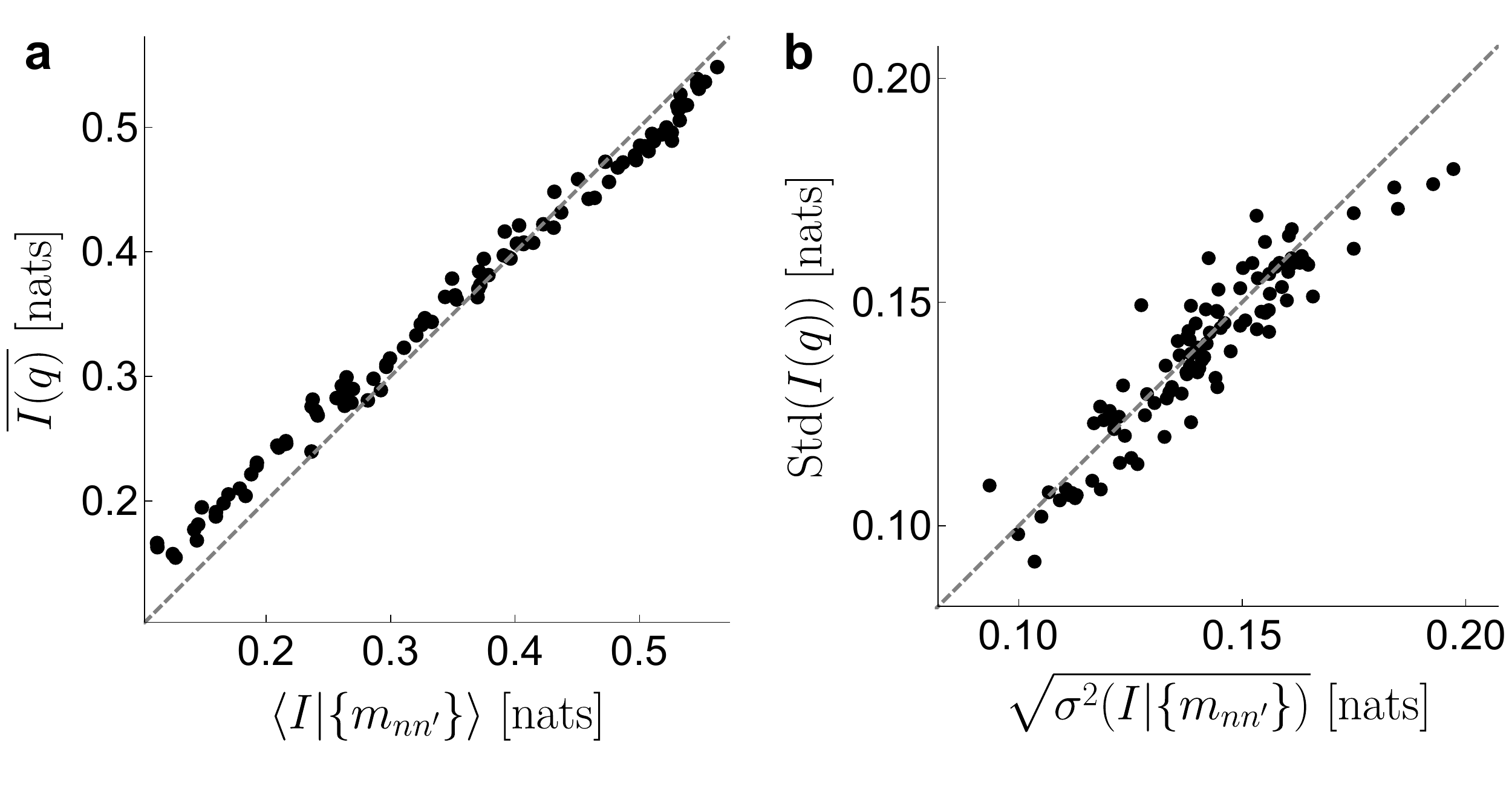}
  \caption{Verification of the accuracy of the analytically predicted mean posterior information (Eq.~\ref{e06:postinfo}) and variance (Eq.~\ref{e:errorih}) in the severely under-sampled regime. A collection of 13,500 distributions $q_{xy}$ are constructed by sampling $q_x\sim\text{DP}(\alpha)$ and $q_{y|x}\sim\text{Beta}(\beta/2,\beta/2)$, with $\alpha$ varying in the set $\{e^4,e^5,e^6\}$ and $\log\beta$ from Eq.~\ref{e04:priorb}. Each distribution $q_{xy}$ has an associated $I_{XY}(q_{xy})$. From each $q_{xy}$, we take five (5) sets of just $N=40$ samples. {\bf a}: The values of $I(q_{xy})$ are grouped according to the multiplicities $\{m_{nn'}\}$ produced by the samples, averaged together, and depicted as the $y$ component of each data point. The $x$ component is the analytical result of Eq.~\ref{e06:postinfo}, based on the sampled multiplicities. {\bf b}: Same analysis for the standard deviation of the information (the square root of the variance calculated in Eq.~\ref{e:errorih}).}
  \label{f03:n40}
\end{figure}

Finally, we check numerically the accuracy of the analytically predicted mean posterior information (Eq.~\ref{e06:postinfo}) and variance (Eq.~\ref{e:errorih}) in the severely under-sampled regime. The test is performed in a different spirit than the numerical evaluations of Fig.~\ref{f02:est}. There, averages were taken for multiple samples of the vector $\mathbf{n}$, from a fixed choice of the probabilities $\mathbf{q}_x$ and $\mathbf{q}_{y|x}$. The averages of Eqs.~\ref{e06:postinfo} and Eq.~\ref{e:errorih}, however, must be interpreted in the Bayesian sense. The square brackets in $\langle I|\mathbf{n} \rangle$ and $\langle H^2_{Y|X} \rangle$ represent averages taken for a fixed data sample $\mathbf{n}$, and unknown underlying probability distributions $\mathbf{q}_x$ and $\mathbf{q}_{y|x}$. We generate many such distributions with $q_x\sim\text{DP}(\alpha)$ (a Dirichlet Process with concentration parameter $\alpha$) and  $q_{y|x}\sim\text{Beta}(\beta/2,\beta/2)$. A total of $13,500$ distributions $q_{xy}$ are produced, with $\log\beta$ sampled from Eq.~\ref{e04:priorb}, and three equiprobable values of $\alpha=\{e^4,e^5,e^6\}$. For each of these distributions we generate five (5) sets of just $N=40$ samples, thereby constructing a list of $5 \times 13,500$ cases, each case characterized by specific values of $\alpha, \beta, \mathbf{q}_x, \{\mathbf{q}_{y|x}\}, I(\mathbf{q}_x, \{\mathbf{q}_{y|x}\}), \mathbf{n}, \langle I|\mathbf{n} \rangle$ and $\sigma^2(I| \mathbf{n})$ . Following the Bayesian rationale, we partition this list in classes, each class containing all the cases that end up in the same set of multiplicities $\{m_{nn'}\}$ ---for example, $\{m_{10}=36,\,m_{20}=2\}$. For each of the $100$ most occurring sets of multiplicities (which together cover $70\%$ of all the cases), we calculate the mean and the standard deviation of the mutual information $I(\mathbf{q}, \{\mathbf{q}_{y|x}\})$ of the corresponding class, and compare them with our predicted estimates $\langle I|\{m_{nn'}\} \rangle$ and $\langle \sigma^2_I|\{m_{nn'}\} \rangle$, using the prior $p(\log\beta)$ from Eq.~(\ref{e04:priorb}). Figure~\ref{f03:n40} shows a good match between the numerical ($y$-axis) and analytical ($x$-axis) averages that define the mean information (panel {\bf a}) and the standard deviation ({\bf b}). The small departures from the diagonal stem from the fact that the analytical average contains all the possible $\mathbf{q}_x$ and $\{\mathbf{q}_{y|x}\}$, even if some of them are highly improbable for one given set of multiplicities. The numerical average, instead, includes the subset of the 13,500 explored cases that produced the tested multiplicity. All the depicted subsets contained many cases, but still, they remained unavoidably below the infinity covered by the theoretical result.

We have also tested cases where $Y$ takes more than two values, and where the marginal distribution $q_y$ is not uniform, observing similar performance of our estimator.


\section{A prior distribution for the large entropy variable}\label{sec:inferringqx}

The prior considered so far did not model the probability $\mathbf{q}_x$ of the large-entropy variable $X$. Throughout the calculation, the probabilities $\mathbf{q}_x$ were approximated by the maximum likelihood estimator $\hat{q}_X= n_x/N$. Here we justify such procedure by demonstrating that proper Bayesian inference on $\mathbf{q}_x$ hardly modifies the estimation of the mutual information. To that end, we replace the prior of Eq.~\ref{e107:nuestraprior} by another prior that depends on both $\mathbf{q}_x$ and $\{\mathbf{q}_{y|x}\}$.

The simplest hypothesis is to assume that the prior $p(\mathbf{q}_x, \{\mathbf{q}_{y|x}\})$ factorizes as $p(\mathbf{q}_x) \ p(\{\mathbf{q}_{y|x}\})$, implying that the marginal probabilities $q_x$ are independent of the conditional probabilities $q_{y|x}$. We propose $q_x\sim \text{DP}(\alpha)$, so that the marginal probabilities $q_x$ are drawn from a Dirichlet Process with concentration parameter $\alpha$, associated to the total number of pseudo-counts. After integrating in $\mathbf{q}_x$ and in $\mathbf{q}_{y|x}$, the mean posterior mutual information for fixed hyper-parameters $\beta$ and $\alpha$ is
\begin{eqnarray}
\langle I|\mathbf{n} \rangle (\beta,\alpha) &=& \frac{N}{N + \alpha}\left\{\hat{H}(Y)-\sum_{x,\,n_x>0} \frac{n_x}{N} \ \left[\psi_0(\beta + n_x + 1) -\sum_y \frac{\beta \hat{q_y} + n_{xy}}{\beta + n_x} \, \psi_0(\beta \hat{q_y} + n_{xy} + 1)     \right]\right\} \nonumber \\ & &  +\frac{\alpha}{N + \alpha} \ \left[\hat{H}(Y)-\psi_0(\beta + 1) +\sum_y  \hat{q_y}\,  \psi_0(\beta \hat{q_y} + 1)     \right].
\label{e90:postinfok}
\end{eqnarray}
Before including the prior $p(\mathbf{q}_x)$, in the severely undersampled regime the mean posterior information was approximately equal to the prior information evaluated in the best $\beta$ (Eq.~\ref{e18:postinfok}). The new calculation (Eq.~\ref{e90:postinfok}) contains the prior information explicitly, weighted by $\alpha/(N+\alpha)$, that is, the ratio between the number of pseudo-counts from the prior and the total number of counts. Thereby, the role of the non-observed (but still inferred) states is established. 

The independence assumed between $\mathbf{q}_x$ and $\{\mathbf{q}_{y|x}\}$ implies that 
\begin{equation}
p(\mathbf{n} | \alpha, \beta) = p(\mathbf{n}_x | \alpha) \ p(\mathbf{n} | \beta).
\end{equation}
The inference over $\alpha$ coincides with the one of PYM with the tail parameter as $d=0$ \cite{archer2014bayesian}, since
\begin{equation}
\displaystyle p({\bf n}_x|\alpha)\propto \frac{\Gamma(1+\alpha)}{\Gamma(N+\alpha)}\alpha^{k_1-1},
\label{eq91:marglika}
\end{equation}
where $k_1=\sum_{x, n_x>0}1$ is the number of states $x$ with at least one sample. With few coincidences in $x$, $p(\mathbf{n}_x|\alpha)$ develops a peak around a single $\alpha$ value that represents the number of effective states. Compared to the present Bayesian approach, maximum likelihood underestimates the number of effective states (or entropy) in $x$. Since the expected variance of the mutual information decreases with the square root of the number of effective states, the Bayesian variance is reduced with respect to the one of ML. 



\section{Discussion}


In this work we propose a novel estimator for mutual information of discrete variables $X$ and $Y$, which is adequate when $X$ has a much larger number of effective states than $Y$. If this condition does not hold, the performance of the estimator breaks down. We inspire our proposal in the Bayesian framework, in which the core issue can be boiled down to finding an adequate prior. The more the prior is dictated by the data, the less we need to assume from outside. Equation~\ref{eq22:thm1} implies that the mutual information $I(X, Y)$ is the spread of the conditional probabilities of one of the variables (for example, $q_{y|x}$, but the same holds for $q_{x|y}$) around the corresponding marginal ($q_y$ or $q_x$, respectively). This observation inspires the choice of our prior (Eq.~\ref{e107:nuestraprior}), which is designed to capture the same idea, and in addition, to be analytically tractable. We choose to work with an hyper-parameter $\beta$ that regulates the scatter of $q_{y|x}$ around $q_y$, and not the scatter of $p_{x|y}$ around $q_x$, because the asymmetry in the number of available states of the two variables makes the $\beta$ of the first option (and not the second) strongly modulated by the data, by the emergence of a peak in $p(\mathbf{n}|\beta)$. 


Although our proposal is inspired in previous Bayesian studies, the procedure described here is not strictly Bayesian, since our prior (Eq.~\ref{e107:nuestraprior}) requires the knowledge of $\hat{q}_y$, which depends on the sampled data. However, in the limit in which $q_y$ is well sampled, this is a pardonable crime, since $\hat{q}_y$ is defined by a negligible fraction of the measured data. Still, Bayesian purists should employ a two-step procedure to define their priors. First, they should perform Bayesian inference on the center of the Dirichlet distribution of Eq.~\ref{e107:nuestraprior} by maximizing $p(q_y|\mathbf{n})$, and then replace $\hat{q}_y$ in Eq.~\ref{e107:nuestraprior} by the inferred $q_y$. For all practical purposes, however, if the conditions of validity of our method hold, both procedures lead to the same result.


By confining the set or possible priors $p(\{\mathbf{q}_{y|x}\})$ to those generated by Eq.~\ref{e107:nuestraprior} we relinquish all aspiration to model the prior of, say, $q_{y|x = 3}$, in terms of the observed frequencies at $x = 3$. In fact, the preferred $\beta$ value is totally blind to the specific $x$ value of each sampled datum.  Only the {\sl number} of $x$-values containing different counts of each $y$-value matters. Hence, the estimation of mutual information is performed without attempting to infer the specific way the variables $X$ and $Y$ are related, a property named {\sl equitability} \cite{kinney2014equitability}, and that is shared also by other methods \cite{nemenman2004entropy,archer2013bayesian,archer2014bayesian}. Although this fact may be seen as a disadvantage, deriving a functional relation between the variables can actually bias the inference on mutual information \cite{kinney2014equitability}. Moreover, fitting a relation is unreasonable in the severe under-sampled regime, in which not all $x$-states are observed, most sampled $x$-states contain a single count, and few $x$-states contain more than two counts. At least, without a strong assumption about the probability space. In fact, if the space of probabilities of the involved variables has some known structure or smoothness condition, other approaches that estimate information by fitting the relation first may perform well \cite{kolchinsky2017estimating, belghazi2018mine, safaai2018information}. Part of the approach developed here could be extended to continuous variables or spaces with a determined metric. This extension is left for future work.


The main result of the paper, is that our estimator has small bias, even in the severely under-sampled regime. It outperforms other estimators discussed in the literature (at least, when the conditions of validity hold), and by construction, it never produces negative values. More importantly, it even works in cases where the collection of true conditional probabilities $q_{y|x}$ is not contained in the family of priors generated by $p(\mathbf{q}| \beta)$, as demonstrated by the second and third examples of Sect.~\ref{testing}. In these cases, the success of the method relies on the peaked nature of the posterior distribution for $\beta$. Even if the selected $p(\mathbf{q} | \beta)$ provides a poor description of the actual collection of probabilities, the dominant $\beta$ captures the right value of mutual information. This is the sheer instantiation of the equitability property discussed above.

Our method provides also a transparent way to identify the statistics that matter, out of all the measured data. Quite naturally, the $x$ states that have not been sampled provide no evidence in shaping $p(\beta|\mathbf{n})$, as indicated by Eq.~\ref{e108:pbetan}, and only shift the posterior information towards the prior (Eq.~\ref{e90:postinfok}). More interestingly, the $x$ states with just a single count are also irrelevant, both in shaping $p(\beta|\mathbf{n})$ and in modifying the posterior information away from the prior. These states are unable to provide evidence about the existence of either flat or skewed conditional probabilities $q_{y|x}$. Only the states $x$ that have been sampled at least twice contribute to the formation of a peak in $p(\beta|\mathbf{n})$, and in deviating the posterior information away from the prior.


Several fields can benefit from the application of our estimator of mutual information. Examples can be found in neuroscience, when studying whether neural activity (a variable with many possible states) correlates with a few selected stimuli or behavioral responses \cite{strong1998entropy, tang2014millisecond, maidana2018information}, or in genomics, to understand associations between genes (large-entropy variable) and a few specific phenotypes \cite{butte2000discovering}. The method can also shed light into the development of rate-distortion methods to be employed in situations in which only a few samples are available \cite{tishby2000information, still2004many}. The possibility of detecting statistical dependences with only few samples is of key importance, not just for analyzing data sets, but also to understand how living organisms quickly infer dependencies in their environments and adapt accordingly \cite{fairhall2001efficiency}. 

\vspace{6pt} 



\subsubsection*{Funding}
This research was funded by CONICET, CNEA, ANPCyT Ra\'{\i}ces 2016 grant number 1004.

\subsubsection*{Acknowledgments}
We thank Ilya Nemenman for his fruitful comments and discussions.


\appendix
\section{Expected variance for a symmetric, binary $Y$-variable}
\label{variance}

The posterior variance of the mutual information is
\begin{equation}
\sigma^2(I|\mathbf{n}) = \langle (I | \mathbf{n})^2 \rangle - \langle I | \mathbf{n} \rangle^2.
\end{equation}
In the first place, we demonstrate that this quantity is proportional to $k_x^{-1}$, implying that our estimator becomes increasingly accurate as the number of states of the $X$-variable increases. Given that
\begin{equation} \label{e:approximatei}
\langle I| \mathbf{n} \rangle  \approx \hat{H}(Y) - \langle H_{Y|X} \rangle(\mathbf{n}),
\end{equation}
with
\begin{eqnarray}
    H_{Y|X}(\{q_{1|x}\}) &=& -\sum_x \hat{q}_x \ H_{Y|x}(q_{1|x}), \label{e:hyxsuma} \\
    H_{Y|x}(q_{1|x}) &=& -\left[ q_{1|x} \ \log(q_{1|x}) + (1- q_{1|x}) \ \log(1 - q_{1|x}) \right] \nonumber
\end{eqnarray}
it is easy to show that 
\begin{equation} \label{e:errorih}
    \sigma^2(I | \mathbf{n}) \approx \sigma^2(H_{Y|X} | \mathbf{n}) = \langle H^2_{Y|X}(\{q_{1|x}\})\rangle - \langle H_{Y|X}(\{q_{1|x}\})\rangle^2. 
\end{equation}
In other words, if the marginal entropy is well sampled, the variance in the information is mainly due to the variance in the conditional entropy.
In turn, $H_{Y|X}$ is defined as an average of $k_x$ terms (Eq.~\ref{e:hyxsuma}). The independence hypothesis implied in the prior of Eq.~\ref{e107:nuestraprior}, and in the way different $q_{1|x}$ and $q_x$ factor out in $q(\mathbf{n} | \mathbf{q}_x, \{\mathbf{q}_{y|x}\})$ (Eq.~\ref{e105:multinomial}), imply that the different terms of (Eq.~\ref{e:hyxsuma}) are all independent of each other. The average of $k_x$ independent terms has a variance proportional to 1/$k_x$, so the estimator proposed here becomes increasingly accurate as $k_x$ grows. 

We now derive the detailed dependence of $\langle H_{Y|X}^2 \rangle - \langle H_{Y|X} \rangle^2$ on the sampled data $\mathbf{n}$. The mean conditional entropy $\langle H_{Y|X} \rangle$ can be written in terms of $\langle H_{Y|x} \rangle (n_{x1}, n_{x0}, \beta)$, that is, of the entropy of the variable $Y$ for a particular state $x$ with $n_x=n_{x0}+n_{x1}$ counts at fixed $\beta$ 
\begin{eqnarray}
\langle H_{Y|X} \rangle (\mathbf{n}) &=& \int p(\beta|\mathbf{n}) \ {\rm d}\beta \ \sum_{x} \frac{n_x}{N} \ \langle H_{Y|x}\rangle(n_{x0}, n_{x1}, \beta)  \nonumber \\
\langle H_{Y|x} \rangle (n_{x0}, n_{x1}, \beta)  &=& - \left\{\psi_0(\beta + n_x + 1) - \sum_{y \in \{0, 1\}} \left(\frac{\beta /2 + n_{xy}}{\beta + n_x}\right)  \psi_0(\beta/2 + n_{xy} + 1)     \right\}
\label{e06:postentroruido}
\end{eqnarray}
Similarly, for the second moment,
\begin{eqnarray}
\langle H^2_{Y|X} \rangle(\mathbf{n}) &=& \int {\rm d} \beta\, p(\beta|\mathbf{n})\prod_x\int {\rm d} q_{1|x}\, p(q_{1|x}|n_{x0}, n_{x1}, \beta) \left[\sum_{x'} \hat{q}_{x'}\,H_{Y|x'}(q_{1|x'})\right]^2\nonumber \\
&=& \int {\rm d} \beta\, p(\beta|\mathbf{n})\left[\sum_{x\neq x'} \hat{q}_x \hat{q}_{x'}\, \langle H_{Y|x}\rangle(n_{x0}, n_{x1},\beta) \ \langle   H_{Y|x'}\rangle(n_{x0}, n_{x1}, \beta) + \sum_{x} \hat{q}^2_x \, \langle H^2_{Y|x}\rangle(n_{x0}, n_{x1}, \beta)\right]\nonumber \\
&=& \int {\rm d}\beta\, p(\beta|\mathbf{n})\left[\langle H_{Y|X} \rangle^2(\mathbf{n}, \beta) +\sum_x \hat{q}^2_x\, \text{Var}[H_{Y|x}](n_{x0}, n_{x1}, \beta)  \right],
\label{e15:2ndmomentHYdX}
\end{eqnarray}
In turn, $\text{Var}[H_{Y|x}((n_{x0}, n_{x1}, \beta)] = \langle H_{Y|x}^2(n_{x0}, n_{x1}, \beta) \rangle - \langle H_{Y|x}(n_{x0}, n_{x1}, \beta) \rangle^2$, where the first moment $\langle H_{Y|x}\rangle(n_{x0}, n_{x1}, \beta)$ is given in Eq.~\ref{e06:postentroruido}. The second moment is  \cite{archer2014bayesian},
\begin{eqnarray}
\left\langle H^2_{Y|x}\right\rangle(n_{x1}, n_{x0}, \beta)  &=& \int_0^1 {\rm d} q_{1|x} \ p(q_{1|x}|n_{x0}, n_{x1},\beta)\, H^2_{Y|x}(q_{1|x})\nonumber \\ 
&=& \int_0^1 {\rm d} q_{1|x} \ p(q_{1|x}|n_{x0}, n_{x1},\beta)\, \left[ q_{1|x}\log(q_{1|x}) + (1 - q_{1|x})\log(1 - q_{1|x})\right]^2\nonumber \\
&=& \frac{2\left(\frac{\beta}{2}+n_{x0}\right) \ \left(\frac{\beta}{2}+n_{x1}\right)}{(\beta+n_x+1)(\beta+n_x)}F\left(\frac{\beta}{2}+n_{x0},\,\frac{\beta}{2}+n_{x1}\right)+\nonumber \\
& & +\sum_{y\in\{0,1\}} \frac{\left(\frac{\beta}{2}+n_{xy}\right) \ \left(\frac{\beta}{2}+n_{xy}+1\right)}{\left(\beta+n_x+1\right) \ \left(\beta+n_x\right)}G\left(\frac{\beta}{2}+n_{xy},\,\beta+n_x\right).
\label{e13:2ndmomentH}
\end{eqnarray}
In this equation,
\begin{eqnarray}
F(z_0,z_1) &=&  -\psi_1(z_0+z_1+2)+ \prod_{i\in\{0,1\}}\left[\psi_0(z_i+1)-\psi_0(z_0+z_1+2)\right],\nonumber \\
G(z_i,z)&=& \left[\psi_0(z_i+2)-\psi_0(z+2)\right]^2+ \psi_1(z_i+2)-\psi_1(z+2),
\label{e14:2ndmomentHaux}
\end{eqnarray}
and $\psi_1(z)$ is the first polygamma function. 

Replacing the obtained expressions in Eq.~\ref{e:errorih}, the variance of the estimated information is obtained. The two terms of Eq.~(\ref{e15:2ndmomentHYdX}) represent the two sources of uncertainty of the conditional entropy: the uncertainty of $\beta$ (first term), manifested in the width of $p(\beta | \mathbf{n})$, and the uncertainty of the conditional entropies for a fixed $\beta$ (second term), manifested in the width of $p(\{q_{1|x}\} | \beta)$. As the number of samples decreases, the uncertainty in $\beta$ becomes the dominant term. 

Finally, we need to mention that the approximate symbol in Eq.~\ref{e:approximatei} stems from the fact that we are assuming that $H(Y)$ is well approximated by its maximum-likelihood estimator. We are therefore neglecting the error in the marginal entropy $H_Y$, and assuming that the error in the mutual information only stems from the uncertainty in the conditional entropy $H_{Y|X}$ (Eq.~\ref{e:errorih}). This assumption is well justified in the context explored in this paper, that is, when $H(X) \gg H(Y)$.

\bibliographystyle{unsrt}  
\bibliography{ref}  

\end{document}